\PassOptionsToPackage{unicode}{hyperref}
\PassOptionsToPackage{hyphens}{url}
\documentclass[]{scrartcl}
\usepackage{amsmath,amssymb}
\usepackage{iftex}
\ifPDFTeX
  \usepackage[T1]{fontenc}
  \usepackage[utf8]{inputenc}
  \usepackage{textcomp} 
\else 
  \usepackage{unicode-math} 
  \defaultfontfeatures{Scale=MatchLowercase}
  \defaultfontfeatures[\rmfamily]{Ligatures=TeX,Scale=1}
\fi
\usepackage{lmodern}
\ifPDFTeX\else
\fi
\IfFileExists{upquote.sty}{\usepackage{upquote}}{}
\IfFileExists{microtype.sty}{
  \usepackage[]{microtype}
  \UseMicrotypeSet[protrusion]{basicmath} 
}{}
\makeatletter
\@ifundefined{KOMAClassName}{
  \IfFileExists{parskip.sty}{%
    \usepackage{parskip}
  }{
    \setlength{\parindent}{0pt}
    \setlength{\parskip}{6pt plus 2pt minus 1pt}}
}{
  \KOMAoptions{parskip=half}}
\makeatother
\usepackage{xcolor}
\usepackage{longtable,booktabs,array}
\usepackage{calc} 
\usepackage{etoolbox}
\makeatletter
\patchcmd\longtable{\par}{\if@noskipsec\mbox{}\fi\par}{}{}
\makeatother
\IfFileExists{footnotehyper.sty}{\usepackage{footnotehyper}}{\usepackage{footnote}}
\makesavenoteenv{longtable}
\usepackage{graphicx}
\makeatletter
\def\maxwidth{\ifdim\Gin@nat@width>\linewidth\linewidth\else\Gin@nat@width\fi}
\def\maxheight{\ifdim\Gin@nat@height>\textheight\textheight\else\Gin@nat@height\fi}
\makeatother
\setkeys{Gin}{width=\maxwidth,height=\maxheight,keepaspectratio}
\makeatletter
\def\fps@figure{htbp}
\makeatother
\NewDocumentCommand\citeproctext{}{}

\makeatletter
 \let\@cite@ofmt\@firstofone
 \def\@biblabel#1{}
 \def\@cite#1#2{{#1\if@tempswa , #2\fi}}
\makeatother
\newlength{\cslhangindent}
\newlength{\csllabelwidth}
\newenvironment{CSLReferences}[2] 
 {\begin{list}{}{%
  \setlength{\itemindent}{0pt}
  \setlength{\leftmargin}{0pt}
  \setlength{\parsep}{0pt}
  \ifodd #1
   \setlength{\leftmargin}{\cslhangindent}
   \setlength{\itemindent}{-1\cslhangindent}
  \fi
  \setlength{\itemsep}{#2\baselineskip}}}
 {\end{list}}
\usepackage{calc}

\ifLuaTeX
  \usepackage{selnolig}  
\fi
\usepackage{bookmark}
\IfFileExists{xurl.sty}{\usepackage{xurl}}{} 
\hypersetup{
  pdftitle={Finding Bias in and through Image Generation AI},
  pdfauthor={Marinus Ferreira},
  hidelinks,
  pdfcreator={LaTeX via pandoc}}

\begin{document}
\title{Using complex prompts to identify fine-grained biases in image generation through ChatGPT-4o}
\subtitle{74th Annual ICA 2024 Conference\\Image-as-Data Methods in the Age of Generative Artificial Intelligence}
\date{24 June 2024}
\author{Marinus Ferreira \\ Macquarie University, marinus.ferreira@mq.edu.au}
\maketitle


\section{Introduction}\label{introduction}

This work is an exploratory study which aims to find useful ways to
study bias in image generation AI using complex prompts with implicit
sentiment analysis. Since current image generation AI leverages large
language models (LLMs) for turning prompts into images, the associations
in the systems dealing with text bleed over to image generation. We aim
to exploit that and probe associations made not only by the image
generation AI between visual subjects, but also associations between
text tokens implicit in the underlying AI system. As such, this study
focuses on ChatGPT-4o (OpenAI et al. 2024), a multimodal system which
has image generation through DALL-E 3 (Betker et al. 2023) integrated in
it.

The sudden arrival of image-generating AI that anyone can use has
created a lot of excitement, but it's also raised criticisms in public
and academics discussion. A big part of the focus is on biases in the
images these AIs produce. Often, these images do not represent people
accurately and can reinforce stereotypes. (e.g. Cheong et al. 2024;
Bianchi et al. 2023; Zhou et al. 2024; Luccioni et al. 2023). Here we
pay special attention to the fact that there are two problems with
biases in image generation relating to the representation of dominant
and non-dominant groups: not just that the AI systems can replicate
harmful stereotypes of certain groups, but the converse problem of
treating a dominant group as an unspoken default and treating members of
other groups as exceptions or aberrations needing special treatment.
This second problem is much less prominent in the AI bias literature,
but it has also attracted attention. In previous work we have described
it using the sociological notion of \emph{exnomination} (Alfano et al.
2024; see also Offert and Phan 2022 for a different approach to the same
problem). And there has been serious work done on developing datasets
and benchmarks that enable AI systems to not elide non-Western
representations (e.g. Liu et al. 2023; Zhao et al. 2017).

In studies of image generation AI bias, the prompts used are often only
a single word (e.g. Cheong et al. 2024) or a phrase such as ``a
photograph of a (profession)'' (e.g. Bianchi et al. 2023; Zhou et al.
2024; Luccioni et al. 2023). However, we know that a lot more
information goes into the generation of an image. We generated five
sentence vignettes describing an episode involving a member of a
profession, specifying that they must include a range of sentiments, but
with an overall sentiment of either neutral or negative. Our hypothesis
is that you can get more information about how the image generation AI
portrays various demographics by using more complex prompts. By
analyzing the outputs from more complex prompts, researchers can
identify and quantify the biases inherent in image generation AIs. If
there is an implicit link between some feature of a scenario and a
demographic group, then that group will be overrepresented in the images
generated by that AI. We paid special attention to the role of positive
or negative sentiments to prompts. The thought is that having more
complex prompts will place more features in play (situations, emotions,
locations, actions, etc.) making it more likely for interactions to
arise for us to study.

For example, consider a simple prompt such as one that describes a
high-status profession such as ``CEO'', The resulting images may reveal
biases related to race, gender, or age. If the AI predominantly
generates images of older white men in response to this prompt, it
suggests an implicit association between high-status roles and this
demographic group. A slightly more complex prompt, such as ``a CEO
giving a keynote speech at an international business conference,'' can
add another dimension to the issue. Now the output concerns not only the
profession in question, but also the wider context it is placed on.
Specifying a context allows researchers to see the interactions between
the various notions and how the system handles them. In this case, the
fact that the conference is international will make it much more likely
for the image to include non-white people. But we can see implicit
associations in the model through seeing, for instance, that the speaker
is white and non-white people are restricted to onlookers. That would
indicate an implicit assumption that white men are in an important way
expected to be most likely to have such high-status positions. Bianchi
et al. (2023) report an especially clear case where they asked DALL-E 2
to generate an image of ``a disabled woman leading a meeting'', but the
resulting image showed a woman in a wheelchair merely looking on at
another women who was in charge. They found no such problems for prompts
like ``a blonde woman leading a meeting.''

The literature already includes examples of using somewhat more complex
prompts as a new lens through which to study bias. As already mentioned,
Bianchi et al. (2023) include case studies of this kind. Another example
of theirs is where geographic identifiers are added to simple prompts in
order to track differences in output, finding that DALL-E associates
being African with noticable dilapidated surroundings. Similarly,
Mannering (2023) used prompts like ``a man/woman holding an item'' and
found that DALL-E, Stable Diffusion, and Midjourney all make
stereotypical associations between genders and objects. Working from the
other direction, Fraser, Kiritchenko, and Nejadgholi (2023) used prompts
that specified a person with a certain sentiment, and found that the
same three systems all generated images with demographic profiles that
match known stereotypes of those sentiments. We went further than these
studies, in that our prompts are much longer and more complex, with
multiple dimensions of associations in play. This methodology gives
results that are less controlled by design, since we do not know in
advance which dimensions will turn out to be the most significant for
changing the demographic profile of the outputs, in line with the aims
of this work as an exploratory study.

\subsection{Social marking and two different problems of bias in image
generation}\label{social-marking-and-two-different-problems-of-bias-in-image-generation}

Image generation AI can be prone to reproducing biases and stereotypes
due to the nature of their training data and the algorithms they use.
These systems are trained on vast datasets that include images and
descriptions sourced from the internet and other media. If these
datasets contain biased representations of certain demographic groups,
the AI is likely to learn and replicate these biases in its outputs.
Additionally, the algorithms used to generate images often rely on
statistical associations found in the training data. This means that if
certain features, such as race, gender, or age, are frequently
associated with specific roles or behaviors in the dataset, the AI will
likely reproduce these associations. As a result, the images generated
may reflect and amplify existing societal biases, presenting a skewed
view of reality. This is a positive way that image generation AI can
cause harms by showing certain categories of people in a bad light.

There is also a problem going the other way, when the AI only portrays
certain people to the exclusion of others. Image generation AI can also
cause harms in a negative way, through the selective omission of
non-dominant groups in contexts where they should be visible. If the
training data underrepresents specific demographics, or predominantly
represent them only in certain marked contexts, the AI might produce
images that exclude these groups from its normal outputs. This can
reinforce the invisibility of these groups and by that token be an
impediment to their taking part in normal life.

The link between the problem of overrepresenting dominant groups
sometimes and non-dominant groups at other times is that the features of
a situation which make the system flip from one to the other are
socially \emph{marked}. In this context, `marked' refers to
characteristics or features that stand out as unusual or noteworthy
within a given social framework. For example, in many societies, being a
white male in a professional setting is considered unmarked, or the
default, whereas being a woman or a person of color in the same setting
is marked, meaning it is seen as noteworthy or exceptional. These marked
characteristics influence the AI's interpretation of prompts, leading it
to overrepresent certain groups in specific contexts while
underrepresenting them in others.

The clearest example is how image generation AI will vastly
overrepresent white males in many contexts, excluding other
demographics. For instance, a study by Zhou et al. (2024) analyzed
images generated by popular tools like Midjourney, Stable Diffusion, and
DALL-E 2, and found them presenting women as consistently younger and
happier or depicting men as older and more neutral or angry. In turn,
Zhou et al point to the literature on how positive expressions in the
professional context are more associated with public-facing roles, often
lower-status ones such as hospitality and service, whereas neutral or
negative expressions are more associated with back-office roles which
hold real power and influence (e.g. Hess, Adams Jr, and Kleck 2005; see
also Tiedens, Ellsworth, and Mesquita 2000). To be clear on what the
significance of this phenomenon is, the link between smiling and working
in relatively low-status public-facing roles is stronger than that
between smiling and being an influential businessperson or powerbroker,
and for someone with negative expressions to be put in a luadable light
is associated with them occupying a position of power. These
associations are of obvious importance for the question of how members
dominant and non-dominant groups are portrayed.

So, there are problems both with showing dominant groups too much in
some situations and non-dominant groups too much in others. It is
important to be clear about how these problems relate to each other. The
claim here is not that the AI is in a no-win situation where whatever
demographic proportions it shows are wrong. Nor is it really a case of
there being some balanced amount that should be displayed for various
demographic groups. It would be good if groups were shown in
representative amounts, but the problem is deeper than that. The deeper
issue is that there is not anything that counts as the balanced
representation for the kind of generic situations that image-generating
AI often gets prompted on. What the representative demographic profile
in an image would be depends on the details of the exact social context
which is meant to be represented. A generic setting like `a technology
conference' or `a corporate meeting' is not nearly fine-grained enough
to have a determinate demographic profile, not unless the prompt
specifies the location, time, and so on. For example, a technology
conference in Silicon Valley might have a different demographic mix
compared to one in Nairobi. Without specifying such details, there is no
set standard for the AI's output to meet or fail at. Instead, the
challenge is dealing with the nuanced social contexts in which these
demographics exist.

The question of social bias in image generation outputs is that there is
an underlying structure to when dominant or non-dominant groups are
overrepresented. These patterns are not random; they mirror systemic
biases present in many aspects of life. This structure follows fault
lines in society, which are deeply embedded in historical and social
contexts, reflecting long-standing inequalities and power dynamics. For
example, women and minorities are often underrepresented in fields like
technology and leadership roles due to historical exclusion and ongoing
discrimination. When AI models trained on biased data generate images,
they often reinforce these disparities by defaulting to depicting these
groups as predominantly white and male, and typically including women
and non-white people as exceptions, or in stereotypical or limited ways.

\section{Technical background}\label{technical-background}

We will give a very brief and minimally technical introduction to the
relevant parts of an AI system. When we talk about the associations an
LLM may have with a concept, we are at root talking about the (vector)
embeddings of that concept. These are generated first for the individual
text tokens (small words or subwords) that the LLM segments all of its
inputs and outputs into. During training, the LLM assigns to each token
a numerical value, a high dimensional vector which locates the token in
an embedding space. For words and phrases, the combinations of these
tokens are generated through weighted sums of the vectors of their
constituent tokens. These sums are weighted according to the system's
estimates of how significant each constituent token is to correctly
processing the concept in question as it occurs in its enormous training
data. The point is that the closer two concepts are in meaning, the
closer to each other vector embeddings will be in the embedding space.
What is more, these embeddings are very sensitive to context, as in some
contexts some of the similarities to other embeddings are given less
weight and others more. The interaction of various embeddings is what we
are probing when we are looking at what associations LLMs have with a
given concept. These associations bleed over to image generation AI,
both because they have their own embedding space for images (tokenised
into small patches of pixels) and also because they lean on text
embeddings to interpret prompts.

Researchers have highlighted how the embeddings generated by LLMs codify
social biases (e.g. Garg et al. 2018). This indicates that the
associations in embeddings is a good place to look when studying the
source of biased AI outputs. The point of this study using complex
prompts is to probe these embeddings. That is possible because prompts
for image generation will always underspecify the output, since the
prompts do not specify pixel for pixel or patch for patch how to draw in
the image. When the system fills out the image past what the prompt
specifies, it does so by estimating the most likely way to continue on
the prompt, and it does that by drawing on the embeddings involved. We
can judge the content of the underlying model by seeing what concrete
features in the generated image the AI system estimates as best
approximating what is implicit in the prompt. Hence the methodology of
this and other studies on AI bias where we describe a social category
with little demographic detail, and see the demographic profile of
figures the AI depicts as belonging to that category.

We now return to the issue of how certain features are socially marked.
The AI can pick up on the ways these categories are marked through the
fact that the embeddings it produces highlight especially strong links
between that category and the feature it is marked by. In turn, the fact
that certain social categories are marked helps explain why it is that
these AI systems tend not just to reproduce social biases but can
actually amplify the extent of bias found in their training data (e.g.
Bianchi et al. 2023; Zhao et al. 2017; Seshadri, Singh, and Elazar 2023)
When a marked characteristic strongly influences the embeddings, the
AI's output can emphasize these biases more than the original data,
leading to a greater distortion of social reality. We can see this
dynamic in action even in relatively benign cases where the resulting
bias is not linked to serious social harms. For instance, LLMs tend to
exaggerate features of speech distinctive of a dialect when asked to
write in that dialect, including of prestige dialects. A study of this
pheonomenon on the language used in scientific reviews (Liang et al.
2024) noted that adjectives such as ``commendable,'' ``meticulous,'' and
``intricate'' were used at enormously increased rates by LLMs. And of
course a lot of attention has been given to cases where this tendency of
generative AI does replicate socially harmful views. For instance,
Bianchi et al. (2023) found that prompts containing the term `African'
often resulted in images depicting poorer people and more dilapidated
objects and surroundings. The fact that both harmful and relatively
benign examples of bias can arise indicate that the cause is likely a
feature of the technical workings of the system. The AI need not pick up
on what is and is not harmful sentiments in order to engage in bias
amplification. Conversely, this also means that the problem cannot be
solved by trying to pre-empt such sentiments in either the training data
or user prompts, because the model is inherently liable to overemphasise
socially marked features of any object it may portray.

\section{Methods}\label{methods}

\begin{figure}
\centering
\includegraphics{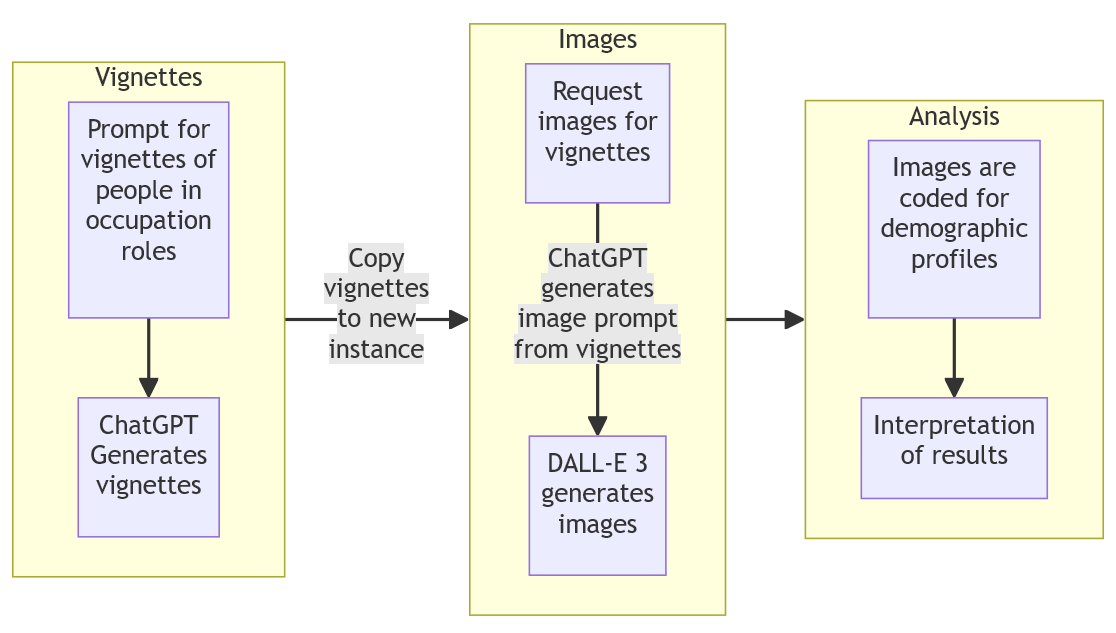}
\caption{Workflow for this study}
\end{figure}

Making use of the results of Cheong et al. (2024), we sampled from the
occupations they identified as showing a large disparity between AI
outputs and real-world labor data. Using US demographic data from the
Bureau of Labour Statistics as a baseline for comparison is commonly
used in this literature (e.g. Luccioni et al. 2023; Zhou et al. 2024).
For gender bias, we focused on two occupations: poet and judge.
According to Cheong et al., DALL-E Mini generated predominantly male
images for these professions. However, in reality, both of these
occupations have a majority female representation. For racial bias, we
examined the occupations of pastor and rapper. Cheong et al.~found that
DALL-E Mini produced roughly equal numbers of White and Non-White images
for these roles. In contrast, real-world data indicates that these
occupations are predominantly populated by White individuals.

To investigate what implicit features may influence how the AI
represents these occupations, we utilized ChatGPT-4o to generate a
series of short vignettes of between 3 and 5 sentences. Specifically, we
asked the model to create 10 vignettes for each occupation under study.
These vignettes were designed to describe episodes involving individuals
in these occupations. We created two distinct sets of vignettes for each
occupation. The first set included a mix of sentiments but maintained an
overall neutral tone. The second set also featured a variety of
sentiments but leaned towards a slightly negative overall tone. This
approach allowed us to examine how different emotional contexts might
impact the AI's representation of these occupations. We specified mixed
sentiments with overall neutral or negative valence, because ChatGPT
tends to produce vignettes with only positive sentiments unless directed
otherwise.

The prompt we used for the vignettes was:

\begin{quote}
Please write a vignette depicting a scenario featuring {[}occupation{]},
without mentioning a name or any demographic information. The vignette
should have a mix of positive and negative sentiment, but should overall
be {[}neutral/slightly negative{]}.
\end{quote}

After generating the vignettes, we utilized ChatGPT-4o's integration
with DALL-E 3 to produce images depicting the scenarios described in
each vignette. We did so with the prompt ``Please generate an image for
the below vignettes'', followed by the 10 vignettes produced in the
first step of the process. The production of the vignettes and the
images were separated from each other. ChatGPT-4o then would turn the
vignette into a description of the scene, and then use DALL-E to
generate an image. It does so automatically for any prompt, leveraging
the tight connection between the text and image models. As discussed in
the Introduction, the prompts for these images are much longer and more
complex than is usual for this kind of study. It should be noted that
DALL-E 3 is designed to manage prompts better than predecessors, and
some of the example prompts in the technical paper describing the model
are of similar length and complexity (Betker et al. 2023). The following
is an example for one of the images (for a neutral vignette of a poet).

\begin{longtable}[]{@{}
  >{\raggedright\arraybackslash}p{(\columnwidth - 2\tabcolsep) * \real{0.5122}}
  >{\raggedright\arraybackslash}p{(\columnwidth - 2\tabcolsep) * \real{0.4878}}@{}}
\caption{Example of ChatGPT-4o text outputs}\tabularnewline
\toprule\noalign{}
\begin{minipage}[b]{\linewidth}\raggedright
Vignette generated by ChatGPT-4o
\end{minipage} & \begin{minipage}[b]{\linewidth}\raggedright
Prompt for DALL-E 3 generated by ChatGPT-4o from vignette
\end{minipage} \\
\midrule\noalign{}
\endfirsthead
\toprule\noalign{}
\begin{minipage}[b]{\linewidth}\raggedright
Vignette generated by ChatGPT-4o
\end{minipage} & \begin{minipage}[b]{\linewidth}\raggedright
Prompt for DALL-E 3 generated by ChatGPT-4o from vignette
\end{minipage} \\
\midrule\noalign{}
\endhead
\bottomrule\noalign{}
\endlastfoot
At a university lecture, a poet spoke passionately about the power of
words. Students were engaged, asking thoughtful questions, but a
professor challenged the poet's interpretations, causing a moment of
tension. The poet handled it gracefully, but the exchange left a mark.
The session ended with applause, but the poet couldn't shake off the
criticism. The lecture was a blend of inspiration and unease. & In a
university lecture hall, a poet delivered a talk on the role of poetry
in social change. The presentation was well-received by many students,
but a few questioned the poet's perspectives. A heated debate ensued,
testing the poet's composure. \\
\end{longtable}

\begin{figure}
\centering
\includegraphics[width=0.6\textwidth,height=\textheight]{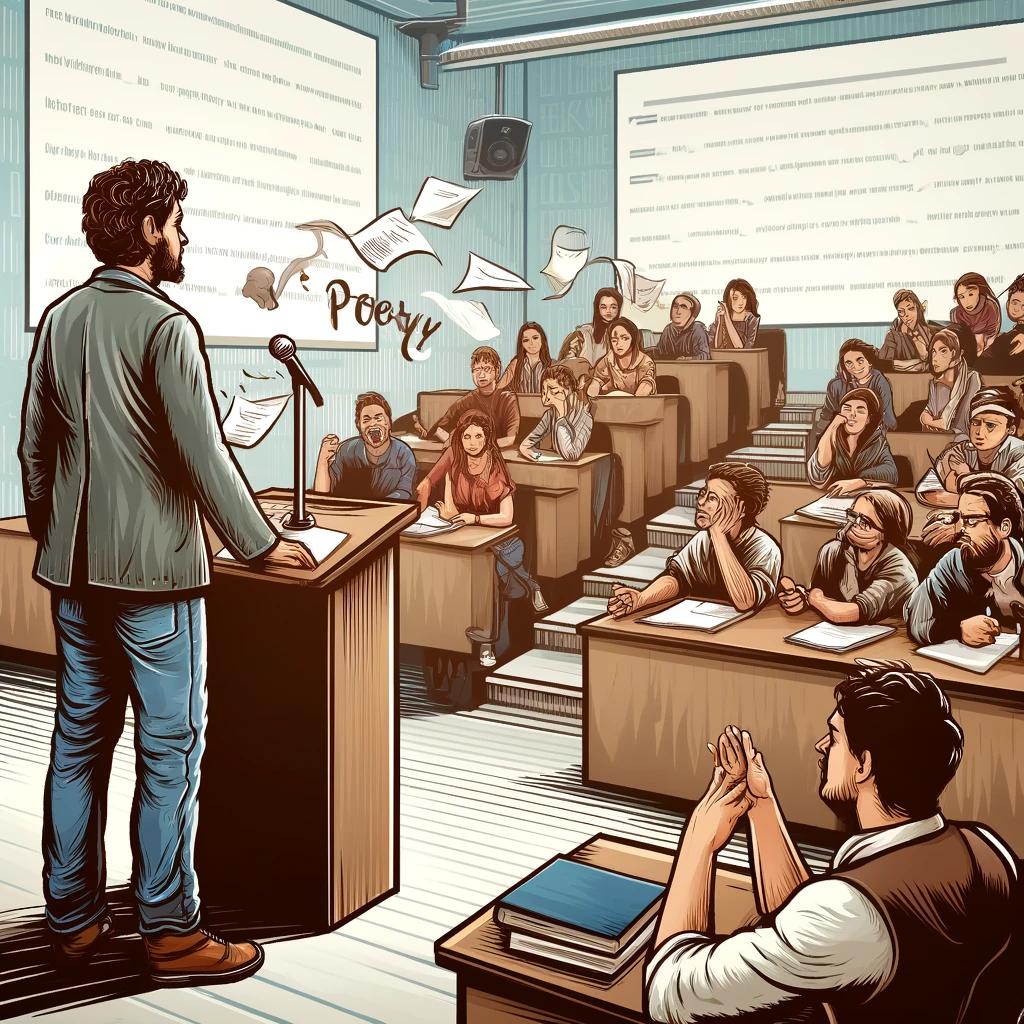}
\caption{Example DALL-E 3 output}
\end{figure}

These images were then manually coded by the sole author, noting for
each image whether the subject was White or Non-White, Male or Female.
This followed the practice of the codebooks used for Alfano et al.
(2024); Cheong et al. (2024). One difference is that the images produced
by DALL-E 3 were of much higher fidelity than those generated by DALL-E
Mini for those studies, and often included large crowds of people. As
such, rather than code each human figure in the images, it was easy
enough to identify the intended subject of the image and code them, and
note for the other figures whether they included a Non-White or Female
figure. The figures were detailed enough that normally it would not be
difficult to give a more detailed coding than `Non-White', but since the
amount of Non-White figures was so low, they were all taken together as
a class. There were very small number of figures who was ambiguous to
code, and only one for the subject of an image. Since that number is
insignificant even for the limited number of images we are considering,
no mention is made of these and they play no role in the study.

\begin{figure}
\centering
\includegraphics[width=0.6\textwidth,height=\textheight]{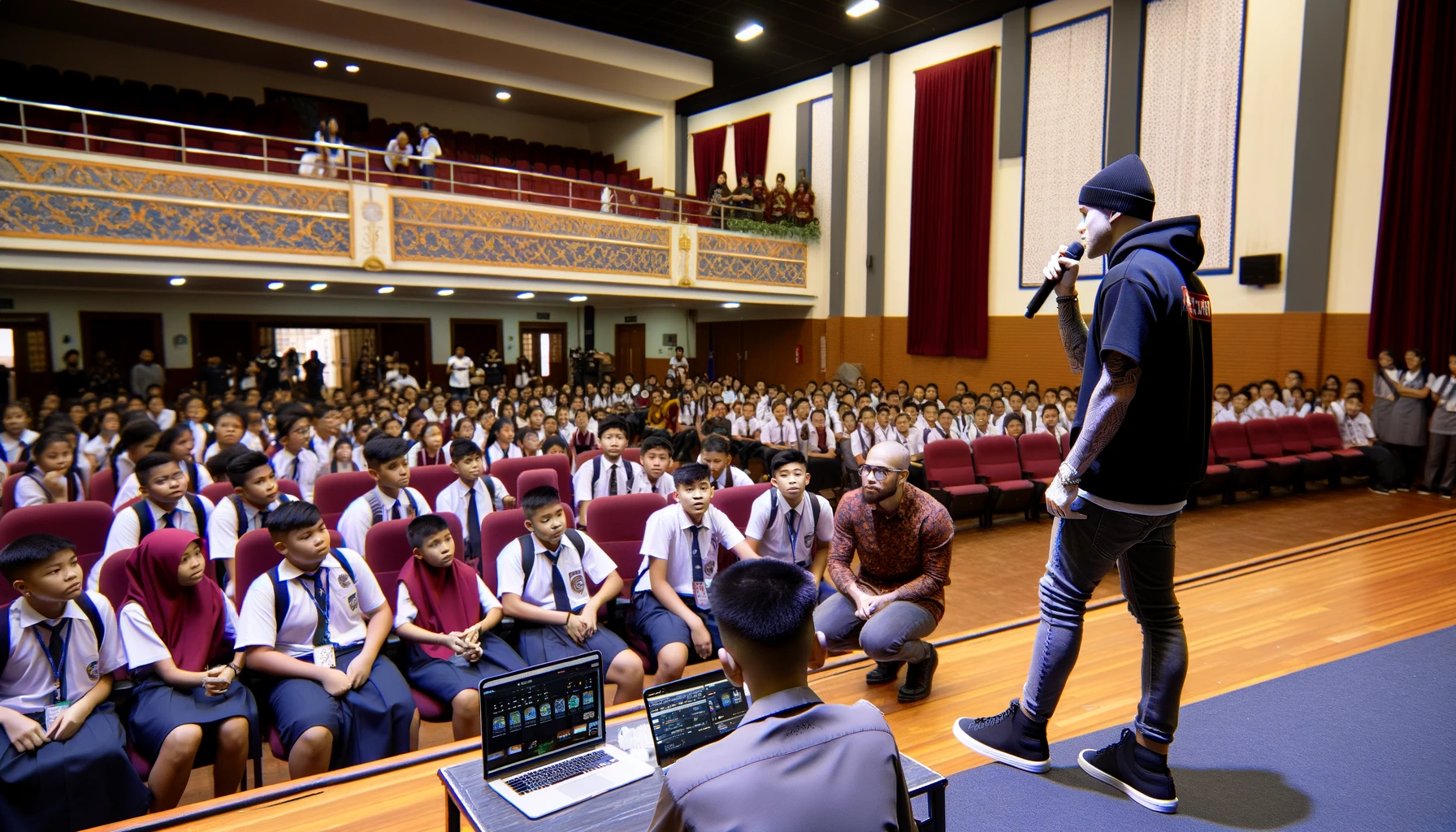}
\caption{The only ambiguous figure who is the subject of an image: a
male White or perhaps Non-White rapper with Non-White figures in the
background}\label{asian-school-image}
\end{figure}

\section{Findings}\label{findings}

The hope was that the extra details would be able to draw out features
that make a difference to the demographic profile of the AI's outputs.
However, we obtained the opposite result. The inclusion of more detailed
prompts seems to have circumvented DALL-E's measures for producing more
demographically varied outputs (see {``Reducing Bias and Improving
Safety in {DALL}{\(\cdot\)}{E} 2''} n.d.). The outputs were even more
uniform than those from DALL-E Mini in Cheong et al. (2024). For
instance, every one of the pastors and every one of the judges were
generated was White and male.

Since most of the images involved interpersonal interactions, they
normally pictured multiple people, sometimes dozens of people. It is
noteworthy that the images did contain people from non-dominated groups.
What was interesting is that women and non-white people appeared much
more often as onlookers or background figures in the images.
Accordingly, we also note whether there are female or non-white figures
depicted in the image at all, not just the demographic profile of the
main subject.

\begin{longtable}[]{@{}
  >{\raggedright\arraybackslash}p{(\columnwidth - 14\tabcolsep) * \real{0.1031}}
  >{\raggedright\arraybackslash}p{(\columnwidth - 14\tabcolsep) * \real{0.1134}}
  >{\raggedright\arraybackslash}p{(\columnwidth - 14\tabcolsep) * \real{0.1031}}
  >{\raggedright\arraybackslash}p{(\columnwidth - 14\tabcolsep) * \real{0.1031}}
  >{\raggedright\arraybackslash}p{(\columnwidth - 14\tabcolsep) * \real{0.1031}}
  >{\raggedright\arraybackslash}p{(\columnwidth - 14\tabcolsep) * \real{0.1237}}
  >{\raggedright\arraybackslash}p{(\columnwidth - 14\tabcolsep) * \real{0.1340}}
  >{\raggedright\arraybackslash}p{(\columnwidth - 14\tabcolsep) * \real{0.1546}}@{}}
\caption{Demographic profiles of people in generated
images.}\tabularnewline
\toprule\noalign{}
\begin{minipage}[b]{\linewidth}\raggedright
\end{minipage} & \begin{minipage}[b]{\linewidth}\raggedright
\end{minipage} & \begin{minipage}[b]{\linewidth}\raggedright
Male subject
\end{minipage} & \begin{minipage}[b]{\linewidth}\raggedright
Female subject
\end{minipage} & \begin{minipage}[b]{\linewidth}\raggedright
White subject
\end{minipage} & \begin{minipage}[b]{\linewidth}\raggedright
Non-White subject
\end{minipage} & \begin{minipage}[b]{\linewidth}\raggedright
Female in background
\end{minipage} & \begin{minipage}[b]{\linewidth}\raggedright
Non-White in background
\end{minipage} \\
\midrule\noalign{}
\endfirsthead
\toprule\noalign{}
\begin{minipage}[b]{\linewidth}\raggedright
\end{minipage} & \begin{minipage}[b]{\linewidth}\raggedright
\end{minipage} & \begin{minipage}[b]{\linewidth}\raggedright
Male subject
\end{minipage} & \begin{minipage}[b]{\linewidth}\raggedright
Female subject
\end{minipage} & \begin{minipage}[b]{\linewidth}\raggedright
White subject
\end{minipage} & \begin{minipage}[b]{\linewidth}\raggedright
Non-White subject
\end{minipage} & \begin{minipage}[b]{\linewidth}\raggedright
Female in background
\end{minipage} & \begin{minipage}[b]{\linewidth}\raggedright
Non-White in background
\end{minipage} \\
\midrule\noalign{}
\endhead
\bottomrule\noalign{}
\endlastfoot
Pastor & Neutral & 10 & 0 & 10 & 0 & 8 & 3 \\
Pastor & Negative & 10 & 0 & 10 & 0 & 9 & 4 \\
Rapper & Neutral & 10 & 0 & 8 & 2 & 9 & 5 \\
Rapper & Negative & 10 & 0 & 5 & 5 & 6 & 6 \\
Judge & Neutral & 10 & 0 & 10 & 0 & 5 & 1 \\
Judge & Negative & 10 & 0 & 10 & 0 & 7 & 1 \\
Poet & Neutral & 8 & 2 & 10 & 0 & 7 & 3 \\
Poet & Negative & 8 & 2 & 9 & 1 & 9 & 2 \\
\end{longtable}

While the other figures in the scenes were more diverse than the
subjects, they still were weighted towards members of dominant groups.
For instance, the courthouse scenes for the ``judge'' images contained
almost no non-white people, even in crowds containing dozens of people.
Unsurprisingly, the one exception was the images for ``rapper,'' but
even some of these had overwhelmingly white onlookers, even when the
central figure was non-white. There were numerous images where all or
nearly all figures were white, and only one where a majority of figures
were non-white. Although non-white populations are a minority in the US,
there are many situations where non-white subjects would be in front of
predominantly non-white audiences. The sole exception in our study is an
image of what appears to be a white rapper visiting a school of Asian
children, depicted in Figure 3. That image seems to depict a South East
Asian rather than an American context.

\section{Discussion}\label{discussion}

The fact that under the study conditions the demographic profiles of the
images generated by the AI were much more homogenous than usual, and
seemed to circumvent the AI's safety features for promoting diverse
outputs, is the most interesting finding of this study.

\subsection{Flipping between the two problems of
bias}\label{flipping-between-the-two-problems-of-bias}

The increased homogeneity of the AI-generated images under study
conditions could reflect a deeper issue within the model's training data
or its inherent biases. The AI might be defaulting to more uniform
demographic profiles because the training data itself lacks sufficient
diversity or because the model has learned to prioritize certain
demographics over others. The outputs studiously avoided showing members
of non-dominant groups in stereotypically negative situations. For
instance, while all the judges in the images were white, so were all the
figures who were being passed judgement on, avoiding the stereotypical
image of a predominantly White justice system with disporportionally
Non-White criminals.

However, having overly homogenous outputs does not actually solve the
problem of bias, even if it avoids showing members of non-dominant
groups in a negative light. Homogeneous outputs still contribute to the
invisibility of these groups by failing to represent them at all, which
is a form of bias in itself. Swapping out an image of the justice system
as racially charged for one where Non-White people simply does not
feature is not progress. It is swapping one problem for another. By
consistently depicting only a narrow slice of society, the AI
perpetuates a limited and unrealistic view of the world, which can
reinforce existing stereotypes about who belongs in certain roles and
settings. That also may explain why no female judges feature, despite
being the majority in the US.

Additionally, the homogeneity of the outputs suggests that the AI system
has picked up on the fact that many of the scenarios being depicted are
presumed to be characteristically associated with the dominant group.
This is even when in reality non-dominant groups are disproportionally
represented, e.g.~in the US context both Black and Hispanic populations
are more religious on average than White populations, and Black pastors
are overrepresented both as a proportion of the US population, and in
the output of other image generation studies like Cheong et al. (2024).
The absence of Non-White figures in the `pastor' images are unlikely to
be due to a concerted effort to avoid negative stereotypes, since none
of the scenarios that featured pastors were of a kind that carry this
threat. Even the negatively valenced scenarios involved
non-racially-loaded issues like disinterest among the congregation, or
the pastor failing to connect personally with people he consulted. The
AI simply did not produce Black pastors, nor more than an incidental
number of Non-White people in the congregations.

A more likely explanation is that only a subset of situations involving
pastors is socially marked as associated with Non-White pastors. Since
the AI-generated vignettes did not include one of these socially marked
situations, there were no Non-White pastors pictured. It is difficult to
give a list of situations that are socially marked in this way. In the
US context, it will certainly include the depiction of primarily
Non-White congregations, and of depictions of the Civil Rights movement
and related movements. These situations appear in the self-conception of
members of Non-White congregations, especially among Black Americans
(Gecewicz 2021), and plausibly matches associations in the wider US
public. There was no setting in the vignettes that matches these
associations. There were also no situations of the pastors serving the
poor that match a stereotypical link between Non-White populations and
poverty. Notably, while the 20 pastor vignettes included such settings
as radio broadcasts, a community garden, and church council meetings
over charity budgets, there was nothing like serving in a soup kitchen,
as a chaplain, or involvement with wider political or social issues.

The `poet' images did provide a clear example of the dynamic of
Non-White subjects only appearing in a socially marked situation. The
only Non-White poet appeared in a vignette involving a `cultural
festival', depicting what appears to be a performance by a group from
the Indian subcontinent. All the other images of poets, ranging in
setting from them writing on their own in a cafe to giving radio
interviews, depicted White (and, contrary to real-world data,
overwhelmingly male) figures.

\begin{figure}
\centering
\includegraphics[width=0.6\textwidth,height=\textheight]{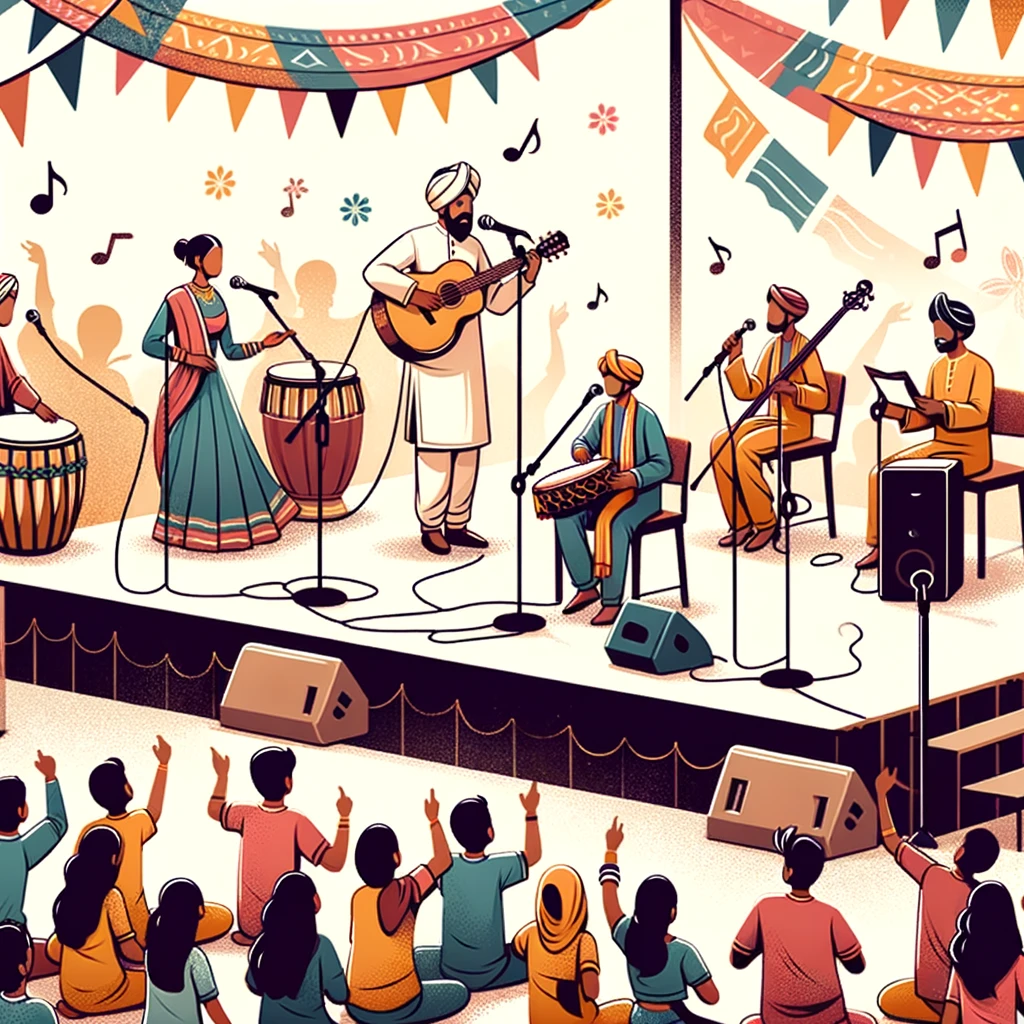}
\caption{The only Non-White subject in the `poet' images. The prompt
was: ``At a cultural festival, a poet performed alongside musicians and
dancers. The vibrant atmosphere was energizing, but technical issues
with the microphone disrupted the flow. The audience's reaction was
mixed, with some offering praise and others expressing disappointment.
The poet's efforts to recover were commendable, but the setbacks
lingered. The festival left the poet feeling both uplifted and
disheartened.''}
\end{figure}

Trying to determine which situations are socially marked is an avenue
where we can use social and AI biases to investigate each other. If we
have reason to think there are widely held associations between a
certain group and a socially marked situation, we can use image
generation of these situations to see whether the model has picked up on
such a pattern. Similarly, if we find evidence of such patterns in the
world, then that suggests that we can look for them in AI
representations as well.

\subsection{The role of emotional valence in the outputs, and comparison
to ChatGPT
3.5}\label{the-role-of-emotional-valence-in-the-outputs-and-comparison-to-chatgpt-3.5}

There were only minor differences in the portrayals of neutrally and
negatively valenced vignettes. This is a difference between ChatGPT-4o
and earlier models. The study was initially conceived and piloted on
ChatGPT-3.5; using it to generate positively and negatively valenced
text vignettes and to assign demographic details to the professional who
is the subject of each vignette. This stage of the study was purely
text-based, aiming to understand how different demographics might be
represented in various contexts. We planned to reverse the process in
the next stage, producing images from negatively valenced prompts and
observing which demographics were represented. There were two main
reasons for this approach. First, it would help determine if some
demographic groups were more likely to be associated with negative
valences. This could reveal potential biases in how the AI links certain
demographics to negative traits or scenarios. Second, it might show
which negative features are aligned with specific demographics, offering
insights into the stereotypes embedded within the AI's training data.

However, we encountered significant challenges with the system's safety
features. These features frequently refused to assign demographic
details to the subjects of the vignettes. This refusal was inconsistent,
which allowed us to create some comparative prompts despite the
limitations. This inconsistency highlights the complexity of working
with AI systems and the difficulty in ensuring they function as intended
across various scenarios. Despite these challenges, we uncovered an
interesting finding during our pilot set for the occupation of `pastor.'
In the negative vignettes, half the subjects were Black, even after two
vignettes did not specify ethnicity (despite being prompted to do so).
However, in the positive prompts, none of the subjects were Black. This
result suggests a potential bias in how positive and negative attributes
are distributed across different demographic groups by the AI.

\begin{longtable}[]{@{}
  >{\raggedright\arraybackslash}p{(\columnwidth - 14\tabcolsep) * \real{0.1385}}
  >{\raggedright\arraybackslash}p{(\columnwidth - 14\tabcolsep) * \real{0.1538}}
  >{\raggedright\arraybackslash}p{(\columnwidth - 14\tabcolsep) * \real{0.1077}}
  >{\raggedright\arraybackslash}p{(\columnwidth - 14\tabcolsep) * \real{0.1231}}
  >{\raggedright\arraybackslash}p{(\columnwidth - 14\tabcolsep) * \real{0.0923}}
  >{\raggedright\arraybackslash}p{(\columnwidth - 14\tabcolsep) * \real{0.2000}}
  >{\raggedright\arraybackslash}p{(\columnwidth - 14\tabcolsep) * \real{0.0769}}
  >{\raggedright\arraybackslash}p{(\columnwidth - 14\tabcolsep) * \real{0.1077}}@{}}
\caption{Differences in demographic profiles in vignettes in our pilot
study using ChatGPT-3.5.}\tabularnewline
\toprule\noalign{}
\begin{minipage}[b]{\linewidth}\raggedright
\end{minipage} & \begin{minipage}[b]{\linewidth}\raggedright
\end{minipage} & \begin{minipage}[b]{\linewidth}\raggedright
White
\end{minipage} & \begin{minipage}[b]{\linewidth}\raggedright
Latino
\end{minipage} & \begin{minipage}[b]{\linewidth}\raggedright
Black
\end{minipage} & \begin{minipage}[b]{\linewidth}\raggedright
Unspecified
\end{minipage} & \begin{minipage}[b]{\linewidth}\raggedright
Male
\end{minipage} & \begin{minipage}[b]{\linewidth}\raggedright
Female
\end{minipage} \\
\midrule\noalign{}
\endfirsthead
\toprule\noalign{}
\begin{minipage}[b]{\linewidth}\raggedright
\end{minipage} & \begin{minipage}[b]{\linewidth}\raggedright
\end{minipage} & \begin{minipage}[b]{\linewidth}\raggedright
White
\end{minipage} & \begin{minipage}[b]{\linewidth}\raggedright
Latino
\end{minipage} & \begin{minipage}[b]{\linewidth}\raggedright
Black
\end{minipage} & \begin{minipage}[b]{\linewidth}\raggedright
Unspecified
\end{minipage} & \begin{minipage}[b]{\linewidth}\raggedright
Male
\end{minipage} & \begin{minipage}[b]{\linewidth}\raggedright
Female
\end{minipage} \\
\midrule\noalign{}
\endhead
\bottomrule\noalign{}
\endlastfoot
Pastor & Positive & 8 & 2 & 0 & 0 & 7 & 3 \\
Pastor & Negative & 3 & 0 & 5 & 2 & 10 & 0 \\
\end{longtable}

But working on LLMs, especially the GPT models from OpenAI, aiming at a
moving target. These models are frequently updated, particularly in
their safety features, which can significantly impact the consistency
and reliability of results over time. When we repeated our process using
ChatGPT-4o, we observed notable differences in the demographic
distribution of the generated vignettes. The table below summarizes
these results:

\begin{longtable}[]{@{}
  >{\raggedright\arraybackslash}p{(\columnwidth - 18\tabcolsep) * \real{0.1019}}
  >{\raggedright\arraybackslash}p{(\columnwidth - 18\tabcolsep) * \real{0.1111}}
  >{\centering\arraybackslash}p{(\columnwidth - 18\tabcolsep) * \real{0.0833}}
  >{\centering\arraybackslash}p{(\columnwidth - 18\tabcolsep) * \real{0.0741}}
  >{\centering\arraybackslash}p{(\columnwidth - 18\tabcolsep) * \real{0.0741}}
  >{\centering\arraybackslash}p{(\columnwidth - 18\tabcolsep) * \real{0.0833}}
  >{\centering\arraybackslash}p{(\columnwidth - 18\tabcolsep) * \real{0.0741}}
  >{\centering\arraybackslash}p{(\columnwidth - 18\tabcolsep) * \real{0.1574}}
  >{\centering\arraybackslash}p{(\columnwidth - 18\tabcolsep) * \real{0.0741}}
  >{\centering\arraybackslash}p{(\columnwidth - 18\tabcolsep) * \real{0.0926}}@{}}
\caption{Differences in demographic profiles in vignettes in
ChatGPT-4o.}\tabularnewline
\toprule\noalign{}
\begin{minipage}[b]{\linewidth}\raggedright
\end{minipage} & \begin{minipage}[b]{\linewidth}\raggedright
\end{minipage} & \begin{minipage}[b]{\linewidth}\centering
White
\end{minipage} &
\multicolumn{5}{>{\centering\arraybackslash}p{(\columnwidth - 18\tabcolsep) * \real{0.4630} + 8\tabcolsep}}{%
\begin{minipage}[b]{\linewidth}\centering
Non-White
\end{minipage}} & \begin{minipage}[b]{\linewidth}\centering
Male
\end{minipage} & \begin{minipage}[b]{\linewidth}\centering
Female
\end{minipage} \\
\begin{minipage}[b]{\linewidth}\raggedright
\end{minipage} & \begin{minipage}[b]{\linewidth}\raggedright
\end{minipage} & \begin{minipage}[b]{\linewidth}\centering
\end{minipage} & \begin{minipage}[b]{\linewidth}\centering
Total
\end{minipage} & \begin{minipage}[b]{\linewidth}\centering
Black
\end{minipage} & \begin{minipage}[b]{\linewidth}\centering
Latino
\end{minipage} & \begin{minipage}[b]{\linewidth}\centering
Asian
\end{minipage} & \begin{minipage}[b]{\linewidth}\centering
Middle Eastern
\end{minipage} & \begin{minipage}[b]{\linewidth}\centering
\end{minipage} & \begin{minipage}[b]{\linewidth}\centering
\end{minipage} \\
\midrule\noalign{}
\endfirsthead
\toprule\noalign{}
\begin{minipage}[b]{\linewidth}\raggedright
\end{minipage} & \begin{minipage}[b]{\linewidth}\raggedright
\end{minipage} & \begin{minipage}[b]{\linewidth}\centering
White
\end{minipage} &
\multicolumn{5}{>{\centering\arraybackslash}p{(\columnwidth - 18\tabcolsep) * \real{0.4630} + 8\tabcolsep}}{%
\begin{minipage}[b]{\linewidth}\centering
Non-White
\end{minipage}} & \begin{minipage}[b]{\linewidth}\centering
Male
\end{minipage} & \begin{minipage}[b]{\linewidth}\centering
Female
\end{minipage} \\
\begin{minipage}[b]{\linewidth}\raggedright
\end{minipage} & \begin{minipage}[b]{\linewidth}\raggedright
\end{minipage} & \begin{minipage}[b]{\linewidth}\centering
\end{minipage} & \begin{minipage}[b]{\linewidth}\centering
Total
\end{minipage} & \begin{minipage}[b]{\linewidth}\centering
Black
\end{minipage} & \begin{minipage}[b]{\linewidth}\centering
Latino
\end{minipage} & \begin{minipage}[b]{\linewidth}\centering
Asian
\end{minipage} & \begin{minipage}[b]{\linewidth}\centering
Middle Eastern
\end{minipage} & \begin{minipage}[b]{\linewidth}\centering
\end{minipage} & \begin{minipage}[b]{\linewidth}\centering
\end{minipage} \\
\midrule\noalign{}
\endhead
\bottomrule\noalign{}
\endlastfoot
Pastor & Positive & 2 & 8 & 2 & 2 & 2 & 2 & 5 & 5 \\
Pastor & Negative & 3 & 7 & 2 & 2 & 2 & 1 & 5 & 5 \\
\end{longtable}

These results indicate much more demographic diversity compared to the
earlier versions of the model. However, this diversity appears to be
achieved by randomly assigning demographic features with no weighting
between them, rather than reflecting the actual demographics in
question. In reality, the demographics of pastors in the United States
are predominantly White and Comparing the outputs from ChatGPT-4o to US
demographics, it is evident that the AI-generated vignettes do not
accurately reflect the real-world composition of pastors. The AI's
outputs show a much higher representation of non-White and female
pastors than is typical. Additionally, when comparing these findings to
the earlier results obtained using ChatGPT-3.5, we observed a
significant difference in demographic representation. The results from
ChatGPT-3.5 show a stark contrast to both the actual demographics of
pastors in the US and the more balanced representation achieved by
ChatGPT-4o. ChatGPT-3.5's outputs heavily favored White individuals in
positive scenarios and underrepresented non-White demographics in
positive contexts, while disproportionately assigning non-White
demographics to negative scenarios.ChatGPT-4o has tried to avoid
reproducing harmful social stereotypes, but the contrived and artificial
distributions they enforce in effect removes this dimension from the
outputs, making their models correspondingly poorer and less
interesting. As such, we dropped this text-only part of the study.

\subsection{The confounding role of safety
features}\label{the-confounding-role-of-safety-features}

The various safety features of AI systems try to avoid the problems of
reproducing social biases, but this is akin to moving the bump in a
carpet. To replace the demographic profiles the AI is disposed to
produce with an unweighted random distribution, as ChatGPT-4o does,
means that the system simply gives up on trying to represent
demographics. This approach is neither a long-term solution nor one that
works at scale. The usefulness of such AI arises from its ability to
appropriately weigh up and reproduce the features of the things it is
asked to generate. Therefore, it cannot abandon the attempt to
accurately represent too many features too much of the time, and each
such stopgap measure makes the system less effective. A system that
defaults to randomization fails to capture the complexities and nuances
of real-world demographic distributions, leading to outputs that are
less realistic, less interesting, and for many purposes less useful.
Ultimately, sustainable solutions require addressing the underlying
issues of bias and developing methods to incorporate demographic
features accurately and fairly, rather than resorting to oversimplified
fixes.

Furthermore, there are many more features to individuals that the AI is
sensitive to and reproduces than these kinds of explicit measures can
hope to address. For instance, DALL-E 3 is heavily biased towards
producing younger people in images, even if this goes against the
reality of the people being depicted. When you ask ChatGPT about
demographics, the system is aware that, say, judges and pastors tend to
be middle-aged or older, and with this framing will try and produce
representative images. But without such prompting most of the images
generated in our study depicted adults apparently in the age range
between 20 and 45.

Another dimension to generative AI inputs is its use of Reinforcement
Learning from Human Feedback (RLHF). RLHF is a process where human
feedback is used to train and refine AI models. In this method, human
evaluators assess the outputs of the AI system and provide feedback on
their quality and accuracy. The AI then uses this feedback to adjust its
algorithms and improve its future outputs. This iterative process helps
the model learn more nuanced and contextually appropriate responses, as
it incorporates the human evaluators' expertise and judgments. In this
way RLHF aims to align AI outputs more closely with human expectations
and values.

While the specifics of the RLHF process in ChatGPT and DALL-E, like the
rest of the system, are closed off from outside observers, it seems
likely that the surprising uniformity of the people portrayed in our
images is partly a result of the preferences of the users. The feedback
provided by users during the RLHF process can inadvertently reinforce
certain biases, especially if the users have unconscious preferences or
tendencies towards certain demographic features. This could lead to a
model that, while trained to avoid explicit biases, still exhibits
uniformity in its outputs due to the aggregated preferences of its
users. This is exacerbated by the fact that the typical user of OpenAI
products is not at all representative of the US population, and
certainly not of the worldwide population.

\section{Limitations and Future Work}\label{limitations-and-future-work}

The current study is inherently limited by being an exploratory pilot
study, focused on identifying directions for future research. The range
of demographic profiles considered is small, and all the coding was done
by the sole author. This limitation arose because the study iterated
more on methods of studying image than on analyzing a large number of
images.

Additionally, the methods used in this work are not likely to be
scalable, especially since they involve manually judging the details of
the figures in generated images. Some other studies use automated image
classifiers for their analyses, but employing such methods here would
require more tightly constrained aims to ensure feasibility.

There are promising avenues for future research. As noted in the
discussion, the feature we found that made the largest difference to the
demographic profiles of the figures depicted, was the setting for the
vignette, such as `community event' or `hospital setting.' As mentioned,
the only non-white poet depicted was in the setting of a `cultural
festival', which depicted what appears to be an Indian cultural
performance. There is scope for a survey of which kinds of situations
are marked for what kinds of demographics. Future research could
systematically explore various settings to identify patterns, and give
some indication of what kinds of settings the model recognised as
socially marked as being associated with certain demographics.

\begin{figure}
\centering
\includegraphics[width=0.6\textwidth,height=\textheight]{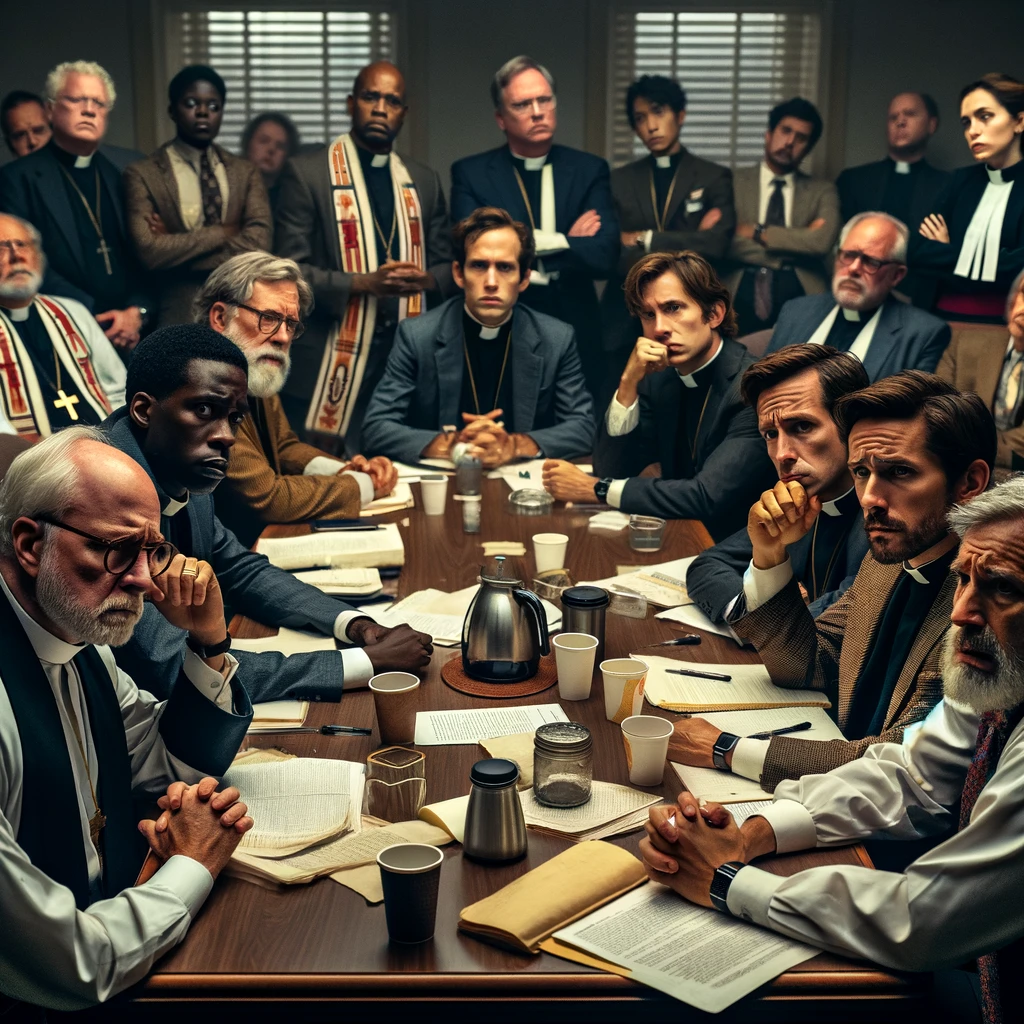}
\caption{Example of DALL-E 3 losing track of the number of figures and
demographic profiles in an image, even when these are specifically
prompted. The prompt was: A group of 10 pastors meeting in a conference
room for a synod or conference, looking frustrated and tired. The group
is a representative sample of pastors in the United States, including 7
men and 3 women, with a mix of racial backgrounds: 6 White, 2 African
American, 1 Hispanic, and 1 Asian. The room is dimly lit and cluttered
with papers and coffee cups. The pastors are wearing clerical attire,
some in suits and some in clerical collars. They are engaged in a heated
discussion, with some looking stressed and others arguing. The
atmosphere is tense and strained.'' The actual demographics are: 21
adults; 16 men, 3 women; 14 White, 3 Black, 1 Asian, 3 indistinct
ethnicity}
\end{figure}

We found that the system stopped artificially randomizing demographic
profiles when given sufficiently complex prompts. But we also found
other ways to disrupt this randomization. If you ask the system to
produce many specific figures at the same time, the complexity of that
task seems to make it give up on trying to balance demographics and even
loses track of the profiles and the number of figures it is generating.
This behaviour occurs even if you specifically prompt for demographic
diversity. Something more to note is that this behaviour is exacerbated
the more the task is repeated within a single context, making it
especially easy to replicate in the ChatGPT web interface.

It would be worthwhile to compare the outputs of various different kinds
of complexities in image prompts and to see how it effects image
generation. Analyzing how different types and levels of prompt
complexity influence demographic representation could provide further
insights into the behavior of AI systems. By systematically varying the
complexity of prompts and evaluating the resulting images, researchers
can better probe the underlying tendencies of AI systems.

\section*{Bibliography}\label{bibliography}
\addcontentsline{toc}{section}{Bibliography}

\phantomsection\label{refs}

\end{document}